\documentclass[prl,aps,twocolumn,reprint,showpacs,floatfix,superscriptaddress,longbibliography]{revtex4-1}
\usepackage[utf8]{inputenc} 
\usepackage{graphicx}
\usepackage{amsmath}
\usepackage{units}
\usepackage{mathtools}
\newcommand{\be}{\begin{equation}}
\newcommand{\ee}{\end{equation}}
\newcommand{\bea}{\begin{eqnarray}}
\newcommand{\eea}{\end{eqnarray}}

\begin{document}
\title{Anisotropic expansion of a thermal dipolar Bose gas}
\author{Y. Tang}
\affiliation{Department of Physics, Stanford University, Stanford CA 94305}
\affiliation{E. L. Ginzton Laboratory, Stanford University, Stanford CA 94305}
\author{A. G. Sykes}
\affiliation{LPTMS, CNRS, Univ.~Paris Sud, Universit\'{e} Paris-Saclay, 91405 Orsay, France}
\author{N. Q. Burdick}
\affiliation{E. L. Ginzton Laboratory, Stanford University, Stanford CA 94305}
\affiliation{Department of Applied Physics, Stanford University, Stanford CA 94305}
\author{J. M. DiSciacca}
\affiliation{E. L. Ginzton Laboratory, Stanford University, Stanford CA 94305}
\affiliation{Department of Applied Physics, Stanford University, Stanford CA 94305}
\author{D. S. Petrov}
\affiliation{LPTMS, CNRS, Univ.~Paris Sud, Universit\'{e} Paris-Saclay, 91405 Orsay, France}
\author{B. L. Lev}
\affiliation{Department of Physics, Stanford University, Stanford CA 94305}
\affiliation{E.~L.~Ginzton Laboratory, Stanford University, Stanford CA 94305}
\affiliation{Department of Applied Physics, Stanford University, Stanford CA 94305}

\date{\today}

\begin{abstract}
We report on the anisotropic expansion of ultracold bosonic dysprosium gases at temperatures above quantum degeneracy and develop a quantitative theory to describe this behavior. The theory expresses the post-expansion aspect ratio in terms of temperature and microscopic collisional properties by incorporating Hartree-Fock mean-field interactions, hydrodynamic effects, and Bose-enhancement factors. Our results  extend the utility of expansion imaging  by providing accurate thermometry for dipolar thermal Bose gases, reducing  error in expansion thermometry from tens of percent to only a few percent. Furthermore, we present a simple method to determine scattering lengths in dipolar gases, including near a Feshbach resonance, through observation of thermal gas expansion.
\end{abstract}

 \maketitle

Expansion imaging of a gas of atoms or molecules after it has been released from a trap provides a simple and highly valuable experimental tool for probing ultracold gases.  For example, the  technique is routinely used for thermometry by measuring the rate of gas expansion as it falls.  The well-established procedure relies on the isotropic expansion of a thermal gas in which the interactions are negligible.   Crucially, deviations from this isotropic behavior can provide a signature of the underlying interactions (and other complex phenomena) within the gas.  Two notable examples of such deviation, caused by interacting systems confined in anisotropic traps, involve an aspect ratio (AR) inversion in non-dipolar Bose-Einstein condensates (BEC) due to mean-field (MF) pressure forces arising from contact interactions~\cite{Anderson1995,Davis1995} and in thermal Bose~\cite{Shvarchuck2003} and degenerate Fermi gases~\cite{Trenkwalder2011} in the collisional-hydrodynamic regime. Both effects alter the time-of-flight (TOF) dynamics and require a  theoretical analysis to be understood~\cite{PedriGueryOdelinStringari2003}. The case of dipolar gases is more complicated since the anisotropy of the interaction also contributes to the TOF AR~\cite{Lahaye:2009kf,Lu2011,Aikawa2014}. No theory  exists for thermal dipolar Bose gas expansion even though such a theory is crucial for accurate thermometry.

In this Letter, we report on the anisotropic expansion of {\it thermal} bosonic $^{162}$Dy and $^{164}$Dy gases~\footnote{Anisotropic expansion of quantum degenerate dipolar Bose and Fermi gases have been explored in Refs.~\cite{Lahaye:2009kf,Lu2011,Aikawa2014}.} and infer the temperature and scattering length from the TOF anisotropy. We find that the dominant physical mechanism responsible for the anisotropy comes from interatomic collisions which partially rethermalize the gas during the TOF.  Non-negligible contributions arise also from Hartree-Fock mean-field interactions and Bose-enhancement factors. In particular, the resulting theory allows us to characterize the background scattering length and width of the 5.1-G Feshbach resonance in $^{162}$Dy~\cite{Baumann:2014ey}. 

Our results pave a way toward investigations of ultracold gases in nontrivial regimes of classical fluid dynamics~\cite{pitaevskii2012physical} where atomic collisions give rise to viscosity and turbulence~\cite{mccomb1992physics,*uriel1995turbulence,*lesieur2008turbulence}. Anisotropic dipolar interactions lead to a magnetoviscosity which has been studied in the context of classical ferrofluids in archetypal situations involving capillary flow~\cite{McTague69,*Hall69,*Shliomis72,*Martsenyuk74}. While quantum ferrofluidity  below condensation temperature $T_c$  has been explored in Cr BECs~\cite{Lahaye:2009kf}, magnetoviscosity of dipolar Bose systems in the intermediate ultracold regime above $T_c$ has yet to be explored. Such a regime is particularly relevant within the context of future progress toward connecting classical~\cite{mccomb1992physics,*uriel1995turbulence,*lesieur2008turbulence} and quantum~\cite{Skrbek11,*Reeves12,*Tsubota14} regimes of turbulence.  It is therefore of fundamental interest that, in contrast to alkali atoms and Cr, this regime is accessible in these ultracold dysprosium gases with unsurpassed magnetic moment $\mu=10\mu_B$ (Bohr magnetons).  

Strongly dipolar lanthanide gases such as Dy and Er have additional complications associated with extremely dense spectra of Feshbach resonances revealed by atom-loss spectroscopy~\cite{Aikawa2012,Baumann:2014ey,Frisch2014,Maier2015}. Such measurements provide the location, $B_0$, of individual resonances and have stimulated statistical studies on their distribution~\cite{Frisch2014,Maier2015Chaotic}. However, atom-loss spectroscopy alone cannot measure the resonance width $\Delta B$~\cite{Chin2010}, the remaining parameter that is required for quantitative control over the scattering length. To obtain $\Delta B$, scattering lengths near a resonance must be measured.  We demonstrate a particularly simple way of doing so by using fits of the thermal-gas AR expansion to our theory; a related technique was demonstrated for dipolar BECs~\cite{Griesmaier06}.

We prepare ultracold gases of $^{162}$Dy and $^{164}$Dy following procedures described in Ref.~\cite{Tang:2015}. In short, we perform laser cooling in two magneto-optical-trap stages, followed by forced evaporative cooling in a crossed optical dipole trap (ODT) formed by two 1064-nm lasers. During the evaporation, the magnetic field is along the $z$-axis (along gravity) and at a Feshbach resonance-free value of $B=1.580(5)$~G~\footnote{Uncertainties are given as standard errors.}. To measure the AR in TOF of the gas, we suddenly turn off the trap and image the gas along the $y$-axis after 16~ms using absorption imaging. We then fit the atomic density to a 2D-Gaussian function to extract the gas size $\sigma_x$ and $\sigma_z$ along $\hat{x}$ and $\hat{z}$~\footnote{$\sigma$ is the standard deviation of the Gaussian profile.}. The gas AR is defined as $\sigma_z/\sigma_x$.

The dipolar thermal Bose gas used in our experiment consists of $N=1.4(1)\times10^5$ atoms for $^{162}$Dy and $1.2(1)\times10^5$ for $^{164}$Dy. The atoms are prepared in the $|{J=8,m_J=-8}\rangle$ ground state. To study the temperature dependence of the AR, we prepare the same number of atoms in the same trap but at different temperatures: First the gas is evaporated close to degeneracy, then the trap depth is increased, and finally we parametrically heat the gas to the desired temperature by modulating the ODT power.  Before releasing the gas for TOF imaging, we let it thermalize in the trap for 1~s, which is much longer than the few-ms thermalization timescale~\cite{Tang2015}. The final trap frequencies are $[\omega_x,\omega_y,\omega_z]=2\pi\times[107(1),49(5),266(1)]$~Hz for both isotopes. We note that this oblate trap geometry, where the confinement is the strongest along the magnetic field orientation $\hat{z}$, is necessary to avoid dipolar mechanical instabilities when evaporating towards $T_c$~\cite{Koch:2008}.

\begin{figure}[t!]
\includegraphics[width=1.\columnwidth]{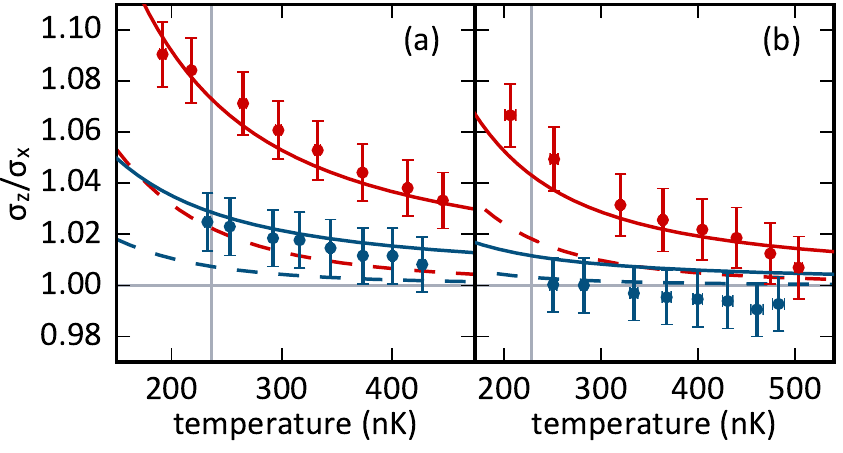}
\caption{Measured gas AR after 16~ms of TOF for $^{162}$Dy in (a) and $^{164}$Dy in (b). In both (a) and (b), red is for magnetic field along $\hat{z}$ and blue is for $\hat{y}$. Points are data with 1$\sigma$ total error: statistical plus 1\% systematic~\cite{Supp}. Solid red and blue curves are calculated using the full theory with the best-fit scattering lengths. Dashed curves are calculated for only the MF effect with the best-fit scattering lengths found using the full theory. Horizontal solid gray line marks unity AR and vertical gray line marks $T_c$.}
\label{T-data}
\end{figure}

The measured gas AR at different temperatures are shown in Fig.~\ref{T-data}. The errors include both statistical and systematic uncertainty and are dominated by systematic error, which we estimate to be 1\%~\cite{Supp}. We measure an anisotropy as large as 9\% for $^{162}$Dy at 200~nK---just below $T_c$---with the field along $\hat{z}$. The anisotropy decreases with higher temperature, or when the magnetic field points along the imaging axis $\hat{y}$, such that the dipolar interaction is symmetric in the imaged $x$-$z$ plane. The same trend is evident for $^{164}$Dy but with overall smaller anisotropy. This field dependence indicates that dipolar physics is at least partially responsible for the anisotropic expansion dynamics, along with the isotope dependence due to different scattering lengths~\cite{Tang2015}, as we now explain.

Our starting point is the known phase-space distribution function of a classical non-interacting gas during expansion $f({\bf r},{\bf p},t)=f_x(x,p_x,t)f_y(y,p_y,t)f_z(z,p_z,t)$, where $f_i(r_i,p_i,t)\propto \exp[-p_i^2/2mk_{\rm B}T-m\omega_i^2(r_i-p_it/m)^2/2k_{\rm B}T]$. The spatial size along direction $i$ evolves according to $\sigma_i(t)=\sqrt{k_{\rm B}T/m\omega_i^2}\sqrt{1+\omega_i^2t^2}$, and in the limit $\omega_i t\gg 1$, we have $\sigma_i(t)\rightarrow \sqrt{k_{\rm B}T/m}t$ leading to the isotropic shape in the long-time limit and reflecting the isotropic momentum distribution in the trap. Even in the presence of interactions, the rapidly decreasing density leads to a saturation of the momentum distribution, with $\sqrt{\langle p_z^2\rangle/\langle p_x^2\rangle}$ determining $\sigma_z/\sigma_x$ after a long TOF. We estimate the finite-$t$ correction to $\sigma_z/\sigma_x$ from the non-interacting case; it scales as $1/t^2$, and for our parameters it does not exceed 0.5\%. Nevertheless, we take this effect into account.

The strategy for calculating $\langle p_i^2\rangle$ relies on a  perturbative treatment.  We write $\langle p_i^2\rangle=mk_{\rm B}T+\Delta\langle p_i^2\rangle$, where $mk_{\rm B}T$ comes from the zeroth-order distribution function $f({\bf r},{\bf p},t)$ and $\Delta\langle p_i^2\rangle$ takes into account interaction and statistical effects. The mean-field (MF) contribution to the kinetic energy $\Delta\langle p_i^2\rangle_{\mathrm{MF}}/2m$ equals work done by the $i$th-component of the gradient of the MF  interaction averaged over $f({\bf r},{\bf p},t)$. This MF part contains the contact term, proportional to the scattering length $a$, and the dipole-dipole term, proportional to the dipole length $a_d=\mu_0\mu^2m/8\pi\hbar^2$~\cite{Bohn2009}, where $\mu_0$ is the vacuum permeability. We find
\begin{equation}
\Delta \langle p_i^2 \rangle_{\mathrm{MF}} =  \frac{2N\hbar^2 \bar{\omega}^3 m^{3/2} }{(k_{\rm B} T)^{3/2}} \{ a_d [H_d^{(i)}+F_d^{(i)}] + a [H^{(i)}+ F^{(i)}]\},
\label{MF}
\end{equation}
where  $\bar{\omega}=(\omega_x\omega_y\omega_z)^{1/3}$ and the dimensionless constants $H$, $H_d$, $F$, and $F_d$ are functions of the trap aspect ratios~\cite{Supp}. These letters stand for the Hartree and Fock contributions, respectively. In addition, the dipole parts $H_d$ and $F_d$ depend on the field orientation \cite{Supp}. Anisotropies due to the MF terms only are shown as dashed lines in Fig.~\ref{T-data}. While the MF interaction is significant, it is not sufficient to match the level of anisotropy observed in our system.

We find that a more important contribution to the AR is the thermalization during the TOF in which the kinetic energy is transferred from $\langle p_i^2\rangle /2m$ to $\langle p_j^2\rangle /2m$ by two-body collisions. In order to understand this phenomenon, we first point to the kinematic effect which occurs in the non-interacting gas and which can be seen from $f({\bf r},{\bf p},t)$: during expansion the thermal motion of particles is transferred to the directed motion characterized by the finite average velocity with components $\langle v_i\rangle=r_i\omega_i^2t/(1+\omega_i^2t^2)$. Important for us is that in the reference frame where the gas is locally stationary, its momentum distribution is equivalent to that of a thermal gas with anisotropic temperature $T/(1+\omega_i^2t^2)$~\cite{Supp}. Collisions try to establish thermal equilibrium by transferring kinetic energy  more frequently, on average, from ``hotter'' directions (smaller $\omega_i$) to ``colder'' ones (larger $\omega_i$). We call this effect hydrodynamic (HD), although the collision rate is too low to continuously maintain thermal equilibrium during expansion. The corresponding contribution to $\Delta \langle p_i^2\rangle$ is linear in the scattering cross section, i.e., quadratic in $a$ and $a_d$,
\bea\label{HD}
 \Delta \langle p_i^2 && \rangle_{\mathrm{HD}} = \\
 &&2 N m^2 a_d^2 \bar{\omega}^2 \Bigg\{ \left[ A_0^{(i)} + A_1^{(i)} \left( \frac{a}{a_d} \right) + A_2^{(i)} \left( \frac{a}{a_d} \right)^2 \right] \nonumber \\
&& + N \left( \frac{\hbar \bar{\omega}}{k_{\rm B} T} \right)^3 \left[ B_0^{(i)} + B_1^{(i)} \left( \frac{a}{a_d} \right) + B_2^{(i)} \left( \frac{a}{a_d} \right)^2 \right] \Bigg\}, \nonumber
\eea
where the dimensionless constants $A$ and $B$ are functions of the trap aspect ratios~\cite{Supp}. The first line in the right hand side of Eq.~\eqref{HD} describes the two-body collisional effects using the differential cross sections obtained in the first-order Born approximation~\cite{BohnJin_PRA_89_022702_2014,Supp}. Previous work on inelastic dipolar collisions has shown the first-order Born approximation to be valid in strongly dipolar systems like dysprosium~\cite{Burdick:2015}. 

The last line in Eq.~(\ref{HD}) accounts for the quantum effects on two-body collisions, where the probability of a scattering event is Bose enhanced according to the local phase-space density. This effect should be distinguished from the deviation of the in situ Bose-Einstein momentum distribution from the Maxwell-Boltzmann one. To first order in the degeneracy parameter, the in situ Bose-Einstein deviation is $\Delta\langle p_i^2 \rangle_{\rm BE} = m k_{\rm B} T (N/16)(\hbar \bar{\omega}/k_{\rm B} T)^3$. It does not introduce any anisotropy to the gas AR, but it is important for the accurate determination of the temperature, even in the non-interacting gas. Adding this correction to the ones given by Eqs.~(\ref{MF}) and (\ref{HD}) results in the {\it corrected} thermometry which infers $T=T_i$ from the expansion dynamics along direction $i$.

Among the four mechanisms labeled by letters $H$ and $F$ in Eq.~\eqref{MF} and $A$ and $B$ in Eq.~\eqref{HD}, we find that the Hartree MF interaction ($H$) and the two-body collision effects ($A$) are the dominant sources of gas anisotropy: For the $^{162}$Dy data point at 200~nK with field along $\hat{z}$ in Fig.~\ref{T-data}(a), they contribute $3.0\%$ and $5.6\%$, respectively, out of the total $9\%$ anisotropy. We have also estimated the effective-range correction to the scattering cross sections by calculating the second-order Born correction to the interaction matrix element at finite collision energy. It is proportional to $a_d^2k$, where $k\propto \sqrt{T}$ is the collision momentum. We find that the corresponding contribution to the AR is negligible for our parameters.

\begin{figure}[t!]
\includegraphics[width=1.\columnwidth]{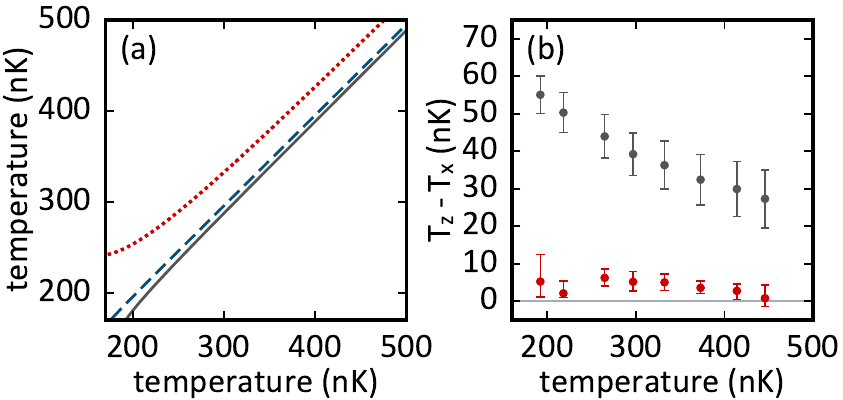}
\caption{(a) Illustration of Bose-corrected TOF thermometry to a dipolar thermal Bose gas, showing that the theory fails to yield the same temperature along the $\hat{x}$, $\hat{y}$, and $\hat{z}$ directions. Field along $\hat{z}$. Theory curves are:  $T_x$ (blue, dashed), $T_y$ (gray, solid), and $T_z$ (red, dotted). (b) Observed difference between $T_x$ and $T_z$, the two dimensions in the imaging plane. The discrepancy is large if only the Bose-corrected TOF thermometry is applied directly (gray points), but can be reduced to close to zero (gray line) using the additional corrections provided in Eqs.~\eqref{MF} and~\eqref{HD} (red points). Theoretical curves in (a) and data in (b) are presented for the experimental parameters used in the $^{162}$Dy measurement of Fig.~\ref{T-data}(a) with the magnetic field along~$\hat{z}$.}
\label{thermometry}
\end{figure}

The MF interaction and the collisional effects cause the gas to expand faster in $\hat{z}$ but slower in $\hat{x}$ and $\hat{y}$ for our system's trap parameters. A direct application of the usual Bose-corrected TOF thermometry (neglecting interactions) in this case would yield conflicting apparent temperatures along each dimension.  Indeed, this is shown by theoretical curves in Fig.~\ref{thermometry}(a). At 200~nK, the discrepancy $\Delta T=T_z-T_x$ between the two dimensions in the imaging plane is about 50~nK, corresponding to 25\% of its temperature.  A mistaken application of this theory leads to an inaccurate determination of temperature and other temperature-related properties such as gas size, trap density, etc., highlighting the need for the corrections in Eqs.~\eqref{MF} and~\eqref{HD}.  

The fact that a gas in thermal equilibrium has a single well-defined temperature allows us to determine the deca-heptuplet $s$-partial-wave scattering length $a$ of $^{162}$Dy and $^{164}$Dy using our theory. With the correct $a$ value, our theory should both minimize $\Delta T$ and predict the measured AR at various temperatures. To determine $a$, we vary $a$ in Eqs.~\eqref{MF} and~\eqref{HD} and find the best-fit scattering length that simultaneously matches the AR data measured at the two different field orientations. In this fitting procedure, we assign the average of $T_x$ and $T_z$ to be the gas temperature. The details of this analysis are described in~\cite{Supp}. The fitted scattering length is $a_{162}=154(22)a_0$ for $^{162}$Dy and $a_{164}=96(22)a_0$ for $^{164}$Dy, where $a_0$ is the Bohr radius.  This new measurement for $^{164}$Dy is consistent with our previously reported value, $92(8)a_0$, measured in cross-dimensional relaxation experiments~\cite{Tang2015}. It also agrees with the measurement reported in Ref.~\cite{Maier2015} using Feshbach spectroscopy. The new best-fit $a$ for $^{162}$Dy is larger than, though not inconsistent with, our previous measurement $122(10)a_0$, and we provide a more detailed discussion of this discrepancy in the supplemental material~\cite{Supp}.

To illustrate that our theory greatly improves the accuracy of thermometry for a thermal dipolar Bose gas, we show in Fig.~\ref{thermometry}(b) $\Delta T$ before and after applying our theory to the $^{162}$Dy measurement. The $\Delta T$ measured in Fig.~\ref{thermometry}(b) increases at lower temperatures and is similar to the theoretical predictions for Bose-corrected TOF thermometry in Fig.~\ref{thermometry}(a). Applying our corrections with the best-fit scattering length leads to almost an order of magnitude reduction in $\Delta T$.  This allows us to determine the temperature of a thermal dipolar Bose gas with far less uncertainty.  The temperatures assigned to the data in Fig.~\ref{T-data} are the average of the corrected $T_x$ and $T_z$; error bars represent the discrepancy.

\begin{figure}[t!]
\includegraphics[width=1.\columnwidth]{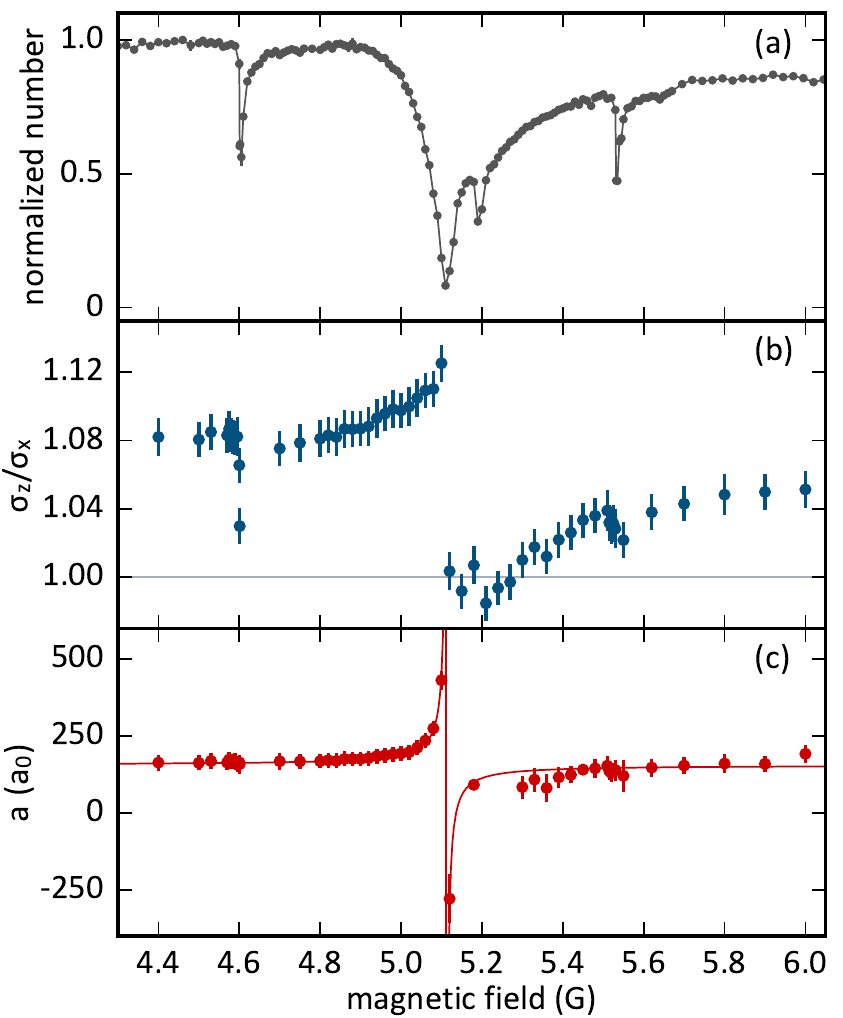}
\caption{(a) High resolution atom-loss spectrum for $^{162}$Dy showing a resonance at 5.1~G and three nearby narrower resonances. Line is guide to eye. (b) Measured gas AR as a function of magnetic field. Horizontal line marks unity AR. (c) Scattering lengths corresponding to the data in (b). We are unable to extract a scattering length for four points near the 5.2~G small resonance with AR below unity; see text for details. All error bars represent 1$\sigma$ uncertainty.}
\label{feshbach}
\end{figure}

The dependence of gas AR on the scattering length $a$ provides an experimental probe for investigating the variation of $a$ near Feshbach resonances. For magnetic Feshbach resonances, $a$ varies with the magnetic field $B$ according to $a(B)=a_{bg}[1-\Delta B/(B-B_0)]$, where $a_{bg}$ is the background scattering length, $B_0$ is the resonance center, and $\Delta B$ is the resonance width~\cite{Chin2010}. We demonstrate the measurement of $a$ near a Feshbach resonance at 5.1~G for $^{162}$Dy, shown in Fig.~\ref{feshbach}(a), by analyzing the gas AR in TOF. Our technique is more convenient than cross-dimensional relaxation for measuring scattering length because it requires only a single experimental measurement to determine $a$ at a given field.  Cross-dimensional relaxation, by contrast, requires multiple measurements to extract a thermalization time as well as extensive numerical simulations when a strong dipolar interaction is present~\cite{SykesBohn_PRA_91_013625_2015}\footnote{An alternative method is to measure the energy gap in the Mott insulating phase at unit filling factor~\cite{Mark2016}.}.

To measure the gas AR near the resonance, we prepare $2.7(1)\times10^5$ atoms at 280~nK in a trap with $[\omega_x,\omega_y,\omega_z]=2\pi\times[89(1),44(5),219(1)]$~Hz.  The magnetic field is first set at 1.580(5)~G, which is the value used for evaporative cooling. We then shift the field to the desired value using a 10-ms linear ramp. Throughout this procedure, the field is kept along the axis of tight confinement, $\hat{z}$, to achieve the largest anisotropy in AR. After the field ramp, we hold the atoms for 50~ms before releasing for TOF imaging.  

The measured gas ARs are shown in Fig.~\ref{feshbach}(b). As the field approaches the 5.1-G resonance from the lower side, we observe increasingly larger AR, as is expected for larger $a$.  We use our theory to convert the AR values to  scattering length, accounting for variations in atom number. The results are shown in Fig.~\ref{feshbach}(c).   The AR that follows from Eqs.~\eqref{MF} and~\eqref{HD} is a quadratic function of $a$ given by $\sigma_z/\sigma_x\approx1.01+(2.3\times10^{-4}+1.6\times10^{-6}\frac{a}{a_0})\frac{a}{a_0}$ for the $\omega$'s, $N$, and $T$ mentioned above. A minimum value therefore occurs at $a\approx-72a_0$ with $\sigma_z/\sigma_x\approx1$. With our 1\% systematic error, we therefore have a {\it blind spot} in scattering length in the region $-139a_0 \lesssim a\lesssim -4a_0$ about $a\approx-72a_0$. It is within this range wherein the four data points near 5.2 G that have ARs below (but within $\sim$$1.5\sigma$ of)  the theoretical minimum value presumably lie, and we are unable to determine a scattering length for them~\cite{blindspot:footnote}. In principle, this blind spot could be shifted to a different region of $a$ by adjusting trap aspect ratios. 

The scattering lengths shown in Fig.~\ref{feshbach}(c) fit well to the functional form $a(B)$. The fitted resonance width is $\Delta B=24(2)$~mG, and the fitted background scattering length is $a_{bg}=157(4)\ a_0$. This $a_{bg}$ value is consistent with the best-fit $a_{162}$ obtained from analysis of the data shown in Fig.~\ref{T-data}(a), which are taken at a different field and trap frequency with about half the atom number. Note that we do not observe a measurable change in $a$ at the other two small resonances near 4.6~G and 5.6~G. 

In conclusion, we observe and develop a theoretical understanding of the anisotropic expansion of \textit{thermal} dipolar Bose gases of $^{162}$Dy and $^{164}$Dy.  The experiment lies in a very favorable regime as far as experiment-theory comparison is concerned; the AR anisotropy is large enough to be measured though small enough for a well-controlled perturbative theory to apply.  As a consequence, we are able to apply this theory for TOF thermometry in this novel regime as well as  measure the scattering length of the gas near a Feshbach resonance with ease.  This simple method for measuring scattering lengths may contribute to the development of a comprehensive theoretical understanding of how collisions are affected within the dense and ultradense Feshbach spectra of these collisionally complex lanthanide atoms~\cite{Frisch2014,Maier2015Chaotic,Maier2015,Burdick:2016vv}. Looking beyond the study of hydrodynamics in magnetic Bose gases, a similar thermometry theory may aid the study of polar molecules near quantum degeneracy.

We acknowledge experimental assistance from Wil Kao, helpful discussions with Matthew Davis, and support from AFOSR, NSF, and the IFRAF Institute. The research leading to these results received funding from the European Research Council (FR7/2007-2013 Grant Agreement No. 341197), and  the European Union's Horizon  2020  research  and  innovation 
programme under grant agreement No 658311.  J.D.~and Y.T.~acknowledge partial support from a Karel Urbanek Postdoctoral Fellowship and a Stanford Graduate Fellowship, respectively.

%

\section{Supplemental Material: \\ Anisotropic expansion of a thermal dipolar Bose gas}

\subsection{Experimental details}\label{exp}

We perform absorption imaging using resonant 421-nm light after 16~ms of time-of-flight (TOF). To image $^{162}$Dy and $^{164}$Dy atoms in their ground state $|{J=8,m_J=-8}\rangle$, we drive $\sigma^-$ transitions to ensure maximal signal-to-noise ratio, which requires a quantization field along $\hat{y}$ because our imaging beam is along $\hat{y}$. For aspect ratio (AR) measurements with the magnetic field along $\hat{z}$, we keep the field along $\hat{z}$ for the first 5~ms of TOF and then rotate to $\hat{y}$ for imaging. We experimentally find that after 5~ms the gas is sufficiently dilute  that the field orientation no longer affects the subsequent expansion dynamics.

We consider the following sources of systematic uncertainties in the AR measurement. (1) Camera alignment with respect to gravity. The angle between camera's $\hat{z}$ axis and gravity is $2.8^{\circ}$, leading to a 0.1\% error. (2) Anisotropy of camera pixels. Data are not available for this error. However, assuming that each pixel's AR varies randomly around unity with a standard deviation at the level of nanofabrication error ($<$10~nm), the anisotropies in the pixel size should average to a negligible amount across our gas size after TOF, which typically consists of about $80\times80$ pixels.  (3) We find the largest systematic error to be the residual inhomogeneity (e.g., interference fringes) in the image beam optical intensity pattern. The fringe structures have a typical length scale of $\sim$10~$\mu$m, which is comparable to our gas size, and have a randomly varying spatial orientation. They  likely  arise from aberrations in various optical elements in the imaging beam path. When we image the same gas at a fixed TOF, but with different part of the imaging beam, the gas AR varies.  This is most likely due to the 2D-Gaussian fit being affected by the fringes in the background. This inhomogeneity in the imaging beam introduces a systematic error of 1\% to AR measurement, which we determine by measuring the AR of a gas with $1.2\times10^5$ $^{164}$Dy atoms at the relatively high temperature of 500~nK and with the field along $\hat{y}$ to reduce the intrinsic AR anisotropy as much as possible.  We repeat the measurement at eight different locations in the imaging beam and take the standard deviation of the eight measurements as this error. 

Since our theory with our trap parameters  predicts allowable $\sigma_z/\sigma_x$ only in the range $\geq$1, the values of $\sigma_z/\sigma_x < 1$ in Fig.~\ref{T-data}(b) may be due to an uncontrolled systematic shift at the $\sim$0.5\% level.  The imperfect nulling of the $T_z - T_x$ difference in Fig.~\ref{thermometry}(b) (red data points) is related to  this systematic shift.

\subsection{Data analysis}\label{analysis}

We fit the atomic density images to a 2D-Gaussian function to extract the gas width $\sigma_x$ and $\sigma_z$:
\begin{equation}
\begin{aligned}
f(x,y) = A e^{ -\frac{(x-x_0)^2}{2\sigma_x^2}} e^{ -\frac{(z-z_0)^2}{2\sigma_z^2}} + m_x x + m_y y + B,
\end{aligned}
\label{2d_gaussian}
\end{equation}
where the linear terms account for the residual gradient in the background and $B$ is the overall offset. The aspect ratio is defined as $\sigma_z/\sigma_x$. We note that the momentum distribution after the expansion deviates from an exact 2D-Gaussian function under the influence of Bose-enhancement, mean-field interaction, and hydrodynamic effects. Nevertheless, we find a 2D-Gaussian fit is sufficiently accurate and robust to extract the second moment of the TOF momentum distribution, allowing for a comparison to the perturbative theory discussed below.
Therefore, the gas width after TOF duration $t$ is given by
\begin{equation}
\begin{aligned}
\sigma_i(t) = \sqrt{\sigma^2_{0,i} + \frac{\langle p^2 \rangle}{m^2} t^2},
\end{aligned}
\label{tof}
\end{equation}
where $\sigma_{0,i}=\sqrt{k_{\rm B} T/m\omega_i^2} $ is the initial trap size along  $i$. 

We use the following procedure to find the best-fit scattering length $a$ from the AR measurements made at different temperatures. First, we estimate a temperature for each AR measurement by numerically solving $T_i$ using the following equation:
\begin{equation}
\begin{aligned}
\sigma_i(t) = \sqrt{\sigma^2_{0,i}(T_i) + \frac{\langle p^2 (T_i,a) \rangle}{m^2} t^2},
\end{aligned}
\label{T-equation}
\end{equation}
where $i=x,z$. We assign the average of $T_x$ and $T_z$ to be the gas temperature. Then we calculate a $\chi^2$ using data taken at both field orientations with the newly assigned temperatures and the theoretical AR values for the corresponding field orientation. The scattering length value $a$ that minimizes $\chi^2$ is the best-fit value. To determine the $1\sigma$ error of $a$, we vary $a$ around its best-fit value in both directions until $\chi^2$ increases by 1~\cite{HughesHayes_MeasurementsAndTheirUncertainties}.

The best-fit $a$ value minimizes the discrepancy between the apparent temperatures $|T_z-T_x|$. We use this fact to determine the scattering lengths for the AR data taken near the 5.1-G Feshbach resonance. For each AR measurement, we numerically find the scattering length $a$ that yields the same $T_x$ and $T_z$. We note that $a$ has two solutions; one must choose the most appropriate solution for each field value. At fields below the resonance center, which is defined as the field with minimum atom number in the high-resolution atom-loss spectrum, we choose the large positive solution as opposed to the large negative solution. From these data points, we estimate the resonance width $\Delta B$, which guides us in choosing the appropriate solution for points immediately above the resonance center according to $a(B)=a_{bg}[1-\Delta B/(B-B_0)]$.  We then fit all $a$ values to this functional form and extract the final fitted $\Delta B$.

\subsection{Scattering length values}\label{scattering_length}

The best-fit scattering length results from this work are $a_{\mathrm{162}}=154(22)a_0$ for $^{162}$Dy and $a_{\mathrm{164}}=96(22)a_0$ for $^{164}$Dy, where $a_0$ is the Bohr radius. While $a_{\mathrm{164}}$ is consistent with our previous measurement from cross-dimensional relaxation experiments, $a_{\mathrm{162}}$ is larger than the previous value $122(10)a_0$~\cite{Tang2015}.  (Note that no value for $a_{\mathrm{162}}$ yet exists from Feshbach data.  Also note the erratum, Ref.~\cite{Tang2016er} below, for Ref.~\cite{Tang2015}.)

\begin{figure}[t!]
\includegraphics[width=1.\columnwidth]{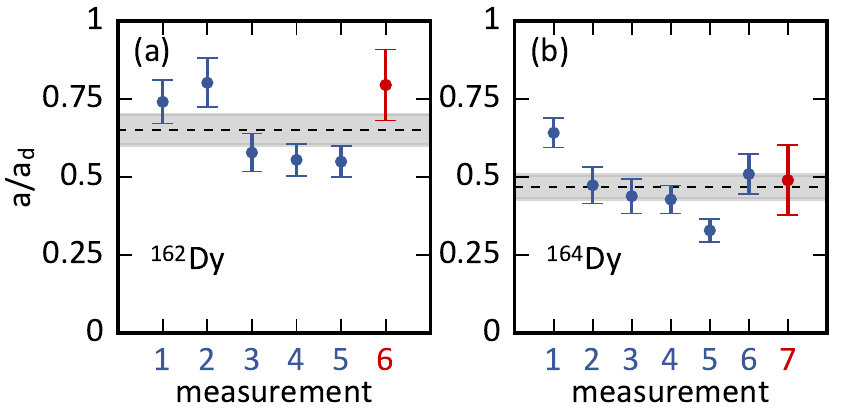}
\caption{Summary of scattering lengths measured from previous cross dimensional relaxation work and our current work for (a) $^{162}$Dy   and (b) $^{164}$Dy. The blue points are from the previous work and the red point is from the current work. The dashed line marks the weighed average and the gray band represents $1\sigma$ error~\cite{Paule:1982}.} 
\label{a_graph}
\end{figure}

We summarize the new and old results in Fig.~\ref{a_graph}. The new result is shown as the red data point.  In our previous cross-dimensional relaxation work, we obtained five independent measurements of scattering lengths for $^{162}$Dy by measuring the rethermalization time at three different field orientations~\cite{Tang2015}. Then we reported the weighted average scattering length~\cite{Paule:1982}. The newly measured $a_{\mathrm{162}}$ is more consistent with measurements 1 and 2 from the previous work, where the field is along $\hat{z}$. The weighted average scattering length value $\bar{a}$, including all measurements, is $\bar{a}_{\text{162}}/a_d=0.65(5)$ for $^{162}$Dy and $\bar{a}_{\text{164}}/a_d=0.47(4)$ for $^{164}$Dy, corresponding to $\bar{a}_{\text{162}}=126(10) a_0$ and $\bar{a}_{\text{164}}=92(8) a_0$; $\bar{a}_{\text{162}}$ ($\bar{a}_{\text{164}}$) is $4a_0$ larger than (same as) the value reported in Ref.~\cite{Tang2015}, and with the same errors.

Using the newly measured scattering lengths, we calculate the Knudsen parameter $\eta$ for trap parameters used for taking the data in Fig.~\ref{T-data}. The Knudsen parameter is defined as the ratio of the mean-free-path $\lambda_i$ to the trap size $l_i$, $\eta = \frac{\lambda_i}{\l_i}$,
where $\lambda_i=(\sqrt{2}n_0\sigma_{\mathrm{tot}})^{-1}$, $l_i=\sqrt{2kT(m\omega_i^2)^{-1}}$, $n_0$ is the peak density, $\sigma_{\mathrm{tot}}$ is the total collision cross section, and $i=x,y,z$~\cite{Shvarchuck2003}. For magnetic atoms, $\sigma_{\mathrm{tot}}$ includes both the $s$-wave collision and the elastic dipolar scattering cross section~\cite{Bohn2009}:
\begin{equation}
\sigma_{\mathrm{tot}} = \sqrt{\left(8 \pi a^2 \right)^2 + \left(2.234 a_d^2 \right)^2}. 
\end{equation}
The criteria for the hydrodynamic regime is $\eta\ll 1$. For our $^{162}$Dy gas at $T=200$ nK, we have $[\eta_x,\eta_y,\eta_z]=[0.37,0.17,0.92]$, and for the $^{164}$Dy gas at the same temperature, we have $[\eta_x,\eta_y,\eta_z]=[0.72,0.33,1.79]$. For both isotopes, the gas lies in the collisionless–hydrodynamic crossover regime.

\subsection{Theory}\label{theory}

{\it {Introduction}:} We begin with the full kinetic equation for the phase-space distribution function of a fluid
\begin{equation}\label{KineticEquation}
    D f(\mathbf{r},\mathbf{p},t)=I[f],
\end{equation}
where 
\begin{equation}\label{D}
D\equiv \frac{\partial}{\partial t}+\frac{\mathbf{p}}{m}\cdot\nabla_{\mathbf{r}}+\mathbf{F}\cdot \nabla_{\mathbf{p}}.
\end{equation}
Interactions  are contained within both the collision functional, $I[f]$, and mean-field forces, $\mathbf{F}$. In our notation, the phase-space distribution is normalized such that $\int\frac{d^3\mathbf{p}}{h^3}f(\mathbf{r},\mathbf{p})=n(\mathbf{r})$, where $n(\mathbf{r})$ is the position space number-density and $h$ is  Planck's constant. To describe the expansion dynamics,  no external (trapping) forces are imposed on the gas (for $t>0$), and $\mathbf{F}$ comes purely from the mean field. That said, the trap will define the initial condition $f(\mathbf{r},\mathbf{p},t=0)$. 

Without interactions, the expansion of the gas is determined by
\begin{equation}
    \partial_t f_0(\mathbf{r},\mathbf{p},t)+\frac{\mathbf{p}}{m}\cdot\nabla_{\mathbf{r}}f_0(\mathbf{r},\mathbf{p},t)=0,
\end{equation}
which has the important general solution $f_0=g_0(\mathbf{p})g_1(\mathbf{r}-\frac{\mathbf{p}}{m}t)$, where $g_0$ and $g_1$ are differentiable functions. 
More directly relevant to a situation where the initial state is that of thermal equilibrium with a temperature well above degeneracy, we have the specific solution
\begin{equation}\label{MBE}
f_{\rm MBE}^{(\mu,T)}=e^{\frac{\mu}{k_{\rm B}T}}\prod_{i=\{x,y,z\}}e^{-\frac{p_i^2/m + m\omega_i^2\left(r_i-p_it/m\right)^2}{2k_{\rm B}T}},
\end{equation}
which corresponds exactly to the Maxwell-Boltzmann initial condition at $t=0$ with chemical potential $\mu$ and temperature $T$.  (The subscript MBE refers to {\it Maxwell-Boltzmann expansion}.)  By integrating over momentum, Eq.~\eqref{MBE} leads to the density distribution expanding as a Gaussian with time dependent standard deviations along each axis $\sigma_i(t)=\sqrt{k_{\rm B}T/m\omega_i^2}\sqrt{1+\omega_i^2t^2}$. With regard to the more intuitive physical description  we present in the main text, we point out that this expansion in Eq.~\eqref{MBE} can be thought of  as describing a {\it wind} which, at a specific location $\mathbf{r}$ and time $t$, {\it blows} with velocity components
\begin{equation}\label{wind}
    v_i\coloneqq\langle p_i/m\rangle=\frac{r_i\omega_i^2t}{1+\omega_i^2t^2},
\end{equation}
where $i=x,y,z$, and the angular brackets denote the {\it momentum-averaged value as a function of position}, defined for instance by
\begin{equation}
\langle \mathbf{p} \rangle=\frac{1}{n(\mathbf{r},t)}\int \frac{d^3\mathbf{p}}{h^3}f(\mathbf{r},\mathbf{p},t)\mathbf{p}.
\end{equation}
Now it is straightforward to calculate the widths (second-moments) of the momentum distribution, with this {\it wind} in Eq.~\eqref{wind} subtracted off, i.e., in the {\it local} rest frame of the gas. This yields
\begin{equation}
 \langle (p_i-mv_i)^2\rangle=\frac{m k_{\rm B}T}{1+\omega_i^2t^2},
\end{equation}
thereby establishing the formal equivalence (within this {\it local} reference frame) to a hypothetical scenario in which a gas has anisotropic temperature $\tilde{T}_i=T/(1+\omega_i^2t^2)$, as discussed in the main text. 

The Maxwell-Boltzmann solution of Eq.~\eqref{MBE} remains useful in a situation where the temperature is approaching degeneracy due to the fact that the Bose-Einstein distribution can be expanded as a series of Maxwell distributions with decreasing temperatures, i.e.,
\begin{equation}\label{BEE}
f_{\rm BEE}^{(\mu,T)}=\sum_{n=1}^\infty f_{\rm MBE}^{(\mu,\frac{T}{n})},
\end{equation} 
where the subscript BEE refers to {\it Bose-Einstein expansion}. The chemical potential $\mu$ in Eqs.~\eqref{MBE} and~\eqref{BEE} is fixed by normalizing the phase-space distribution in the manner mentioned beneath Eq.~\eqref{D}.

{\it {Mean-field interactions}:} To include interactions, we first begin with the mean-field forces. During expansion both Hartree and Fock mean-field potentials contribute to pressure in the gas and their expressions are given by
\bea\label{HartreeFock}
    U_{\rm H}(\mathbf{r})&=&\int d^3\mathbf{r}'\;V_{\rm int}(\mathbf{r}-\mathbf{r}')n(\mathbf{r}')\label{Hartree} \\
    U_{\rm F}(\mathbf{r},\mathbf{p})&=&\int \frac{d^3\mathbf{k}'}{(2\pi)^3}\;\tilde{V}_{\rm int}(\mathbf{k}-\mathbf{k}')f(\mathbf{r},\hbar\mathbf{k}')\label{Fock},
\eea
where wave-number and momenta are related  by $\mathbf{p}=\hbar\mathbf{k}$. The two-body interaction potential $V_{\rm int}(\mathbf{r})$ is given by 
\begin{equation}\label{InteractionPotential}
V_{\rm int}(\mathbf{r})=\frac{4\pi\hbar^2 a}{m}\delta^{(3)}(\mathbf{r})+\frac{2\hbar^2 a_d}{m}\left[
\frac{1-3(\hat{\epsilon}\cdot\hat{r})^2}{r^3}
\right],
\end{equation}
where $a$ is the $s$-wave scattering length and $a_d$ is the dipole length, the $\hat{\phantom{r}}$ denotes a unit vector and $\hat{\epsilon}$ points along the direction of dipole alignment. The Fourier transform of the two-body interaction potential is required in the Fock contribution, Eq.~\eqref{Fock}, and is given by \begin{equation}
\tilde{V}_{\rm int}(\mathbf{k})=\frac{4\pi\hbar^2 a}{m}+\frac{8\pi \hbar^2 a_d}{m}\left[\left(\hat{\epsilon}\cdot\hat{k}\right)^2-\frac{1}{3}\right].
\end{equation}
The force arising from such momentum-dependent mean-field potentials as those in Eqs.~\eqref{HartreeFock} is in general given by
\begin{equation}
\mathbf{F}=m\frac{d\phantom{t}}{dt}\nabla_{\mathbf{p}}U(\mathbf{r},\mathbf{p})-\nabla_{\mathbf{r}}U(\mathbf{r},\mathbf{p}),
\end{equation}
which, after setting $U=U_{\rm H}+U_{\rm F}$, is then inserted into the kinetic equation given in Eq.~\eqref{KineticEquation}.

{\it {Collisions}:} We now consider effects that arise from two-body collisions in the gas which, under the standard assumptions of molecular chaos, can be calculated via the inclusion of the collision integral on the right hand side of Eq.~\eqref{KineticEquation}. This is given by 
\begin{align}\label{CollisionIntegral}
I[f]=\int\frac{d^3 \mathbf{p}_1}{h^3}\int d^2\hat{\Omega}\frac{d\sigma}{d\Omega}v_r\Big[
f'f_1'(1+f)(1+f_1)-\nonumber\\
ff_1(1+f')(1+f_1')
\Big],
\end{align}
where $f=f(\mathbf{r},\mathbf{p})$, $f_1=f(\mathbf{r},\mathbf{p}_1)$, $f'=f(\mathbf{r},\mathbf{p}')$, and $f_1'=f(\mathbf{r},\mathbf{p}_1')$ introduce the four momenta (two incoming and two outgoing) associated with a two-body collision, and $v_r=|\mathbf{p}-\mathbf{p}_1|/m$ is the relative velocity.  It is already assumed in Eq.~\eqref{CollisionIntegral} that these momenta are related by the conservation of energy and momenta, i.e., $\mathbf{p}+\mathbf{p}_1=\mathbf{p}'+\mathbf{p}_1'$ and $p^2+p_1^2=(p')^2+(p_1')^2$, and thus the integration over these additional momenta has been reduced to an integration over just $\mathbf{p}_1$ (the incoming momentum) and $\hat{\Omega}$ (the solid angle through which the relative momentum is rotated during the collision). It is important to note that our expression for the collision integral includes effects due to Bose-enhancement which become increasingly relevant with higher phase-space density (lower temperatures). The differential scattering cross section $\frac{d\sigma}{d\Omega}$ is crucially a function of both incoming and outgoing relative velocities, and in the case of identical bosons scattering at low energy via the interaction potential given in  Eq.~\eqref{InteractionPotential}, this can be calculated in the first-order Born approximation to be~\cite{BohnJin_PRA_89_022702_2014}
\begin{align}
\frac{d\sigma}{d\Omega}\!=\!2a_d^2\Bigg[
\frac{(\hat{p}_r\cdot\hat{\epsilon})^2+(\hat{p}_r'\cdot\hat{\epsilon})^2-2(\hat{p}_r\cdot\hat{\epsilon})(\hat{p}_r'\cdot\hat{\epsilon})(\hat{p}_r\cdot\hat{p}_r')}{1-(\hat{p}_r\cdot\hat{p}_r')^2}\nonumber\\
-\frac{2}{3}+\frac{a}{a_d}
\Bigg]^2,
\end{align}
where $\hat{p}_r$ and $\hat{p}_r'$ denote unit vectors along the direction of relative incoming $(\mathbf{p}-\mathbf{p}_1)$ and outgoing $(\mathbf{p}'-\mathbf{p}_1')$ momentum respectively.

{\it {Equation of change for mean values}:} We are not searching for a full solution to the phase-space distribution, rather just the second moment of the momentum distribution, whose evolution is derived from Eq.~\eqref{KineticEquation} by multiplying by $p_i^2$ (where $i=x,y,z$) and integrating over space and momentum. This moment is what ultimately determines the width of the expanded image after a sufficiently long period of TOF. Let $\chi$ denote the dynamical variable of interest (i.e., $\chi=p_x^2$, say). The equation of change for $\chi$, found by multiplying Eq.~\eqref{KineticEquation} by $\chi$ and then integrating over $\mathbf{p}$, is given by
\begin{equation}\label{Enskogg}
\partial_t\langle n \chi\rangle = n\langle D\chi\rangle-\nabla_{\mathbf{r}}\cdot\langle n\chi\frac{\mathbf{p}}{m}\rangle+\mathcal{C}[\chi],
\end{equation}
where $D$ is defined in Eq.~\eqref{D}. The collisional contribution can be rearranged, using the energy and momentum conservation laws, into the form
\begin{equation}\label{C}
\mathcal{C}(\chi)=\frac{1}{2}\int\! \frac{d^3\mathbf{p}}{h^3}\int\! \frac{d^3\mathbf{p}_1}{h^3}\int\! d^2\hat{\Omega} \frac{d\sigma}{d\Omega}v_r  \Delta\!\chi f f_1 (1+f'+f_1'),
\end{equation}
where $\Delta\!\chi=\chi'+\chi_1'-\chi-\chi_1$, with $\chi=\chi(\mathbf{r},\mathbf{p})$, $\chi_1=\chi(\mathbf{r},\mathbf{p}_1)$, $\chi'=\chi(\mathbf{r},\mathbf{p}')$, and $\chi_1'=\chi(\mathbf{r},\mathbf{p}_1')$. The terms inside the parentheses of Eq.~\eqref{C} arise from the Bose-enhancement factors. We also average over space to find the total average, defined by
\begin{equation}
\langle\!\langle A\rangle\!\rangle=\frac{1}{N}\int d^3\mathbf{r}\langle A\rangle n(\mathbf{r},t)
\end{equation}
where $N$ is the total particle number. This total average is a function of time only. In this way, Eq.~\eqref{Enskogg} leads us to the expression 
\begin{equation}
\partial_t\langle\!\langle \chi\rangle\!\rangle=\langle\!\langle \mathbf{F}\cdot\nabla_{\mathbf{p}}\chi\rangle\!\rangle+\frac{1}{N}\int d^3\mathbf{r}\mathcal{C}[\chi],
\end{equation}
where we have assumed that $\chi$ is not explicitly a function of time or space (recall $\chi=p_i^2$ where $i=x,y,z$). Integrating this ordinary differential equation, and taking the limit $t\rightarrow\infty$, we find
\begin{equation}\label{chi}
\langle\!\langle \chi\rangle\!\rangle_{t\rightarrow\infty}=\int_0^\infty dt\left[
\langle\!\langle \mathbf{F}\cdot\nabla_{\mathbf{p}}\chi\rangle\!\rangle+\frac{1}{N}\int d^3\mathbf{r}\mathcal{C}[\chi]
\right],
\end{equation}
which is all that we require to proceed with our perturbative solution.

{\it {Perturbative solution}:} Our perturbative solution operates under the assumption that the expansion of the gas is dominated by the free-expansion solution given by Eqs.~\eqref{MBE} and \eqref{BEE}. Under this assumption, one can simply plug these formulae into Eq.~\eqref{chi}, and the remaining task of computing all the integrals is straight-forward, albeit arduous. We truncate the sum in Eq.~\eqref{BEE} to $n=2$, thus restricting ourselves to a first-order approximation of the effects due to Bose-Einstein statistics. Accordingly, we expand Eq.~\eqref{C} to first order in the degeneracy parameter $N \left( \hbar \bar{\omega}/k_{\rm B} T\right)^3$. The zeroth order terms in this degeneracy parameter establish the constants $A_{0,1,2}$ of Eq.~\eqref{HD} in the main text, while the first order terms establish the $B_{0,1,2}$.

After some work, we find the second moment of the gas momentum is
\begin{equation}
\begin{aligned}
\langle & p_i^2  \rangle =  m k_{\rm B} T \left[ 1 + \frac{N}{16} \left( \frac{ \hbar \bar{\omega} }{ k_{\rm B} T } \right)^3 \right] \\
+& 2N \frac{\hbar^2 \bar{\omega}^3 m^{3/2} }{(k_{\rm B} T)^{3/2}} \left( a_d H_d^{(i)} + a H^{(i)} \right) \\
+& 2N \frac{\hbar^2 \bar{\omega}^3 m^{3/2} }{(k_{\rm B} T)^{3/2}} \left( a_d F_d^{(i)} + a F^{(i)} \right) \\
+& 2 N m^2 a_d^2 \bar{\omega}^2 \Bigg\{ \left[ A_0^{(i)} + A_1^{(i)} \left( \frac{a}{a_d} \right) + A_2^{(i)} \left( \frac{a}{a_d} \right)^2 \right]  \\
+& N \left( \frac{\hbar \bar{\omega}}{k_{\rm B} T} \right)^3 \left[ B_0^{(i)} + B_1^{(i)} \left( \frac{a}{a_d} \right) + B_2^{(i)} \left( \frac{a}{a_d} \right)^2 \right] \Bigg\}, \\
\end{aligned}
\label{p_square}
\end{equation}
where $i=x,y,z$ denotes the axis, $\bar{\omega} = (\omega_x \omega_y \omega_z)^{1/3}$ is the geometric mean trap frequency, $T$ is temperature, $N$ is atom number, $a_d=\mu_0\mu^2 m/(8\pi\hbar^2)$ is the dipole length scale, and $a$ is the $s$-wave scattering length. The dimensionless constants $H_d$, $H$, $F_d$, $F$, $A$, and $B$ are remnants of integration over the solid angles of incoming and outgoing momenta and are given in the appendix of this supplement. These turn out to be complicated, mainly by the expression for the differential cross section, and the easiest approach is to simply compute these numerically for a given set of trap frequencies.

The first line in Eq.~\eqref{p_square} comes from the expansion of a non-interacting gas, including the Bose statistics, to  first order in the degeneracy parameter. The second and third line are derived from the Hartree and Fock mean-field interactions, respectively. The fourth line accounts for the two-body collisional effects during expansion, and the fifth line describes the Bose-enhancement correction to the collision integral. 

In addition, we have looked at results involving the full summation in Eq.~\eqref{BEE}, i.e., including effects due to the degeneracy parameter $N \left( \hbar \bar{\omega}/k_{\rm B} T\right)^3$ at all orders. In this case, the integrals associated with calculating $\langle p_i^2  \rangle$ become considerably more complicated. However, we were able to compute an upper bound on $\langle p_i^2 \rangle$ by replacing the integrand with an absolute value. We found that even this upper bound contributes negligible difference compared to Eq.~\eqref{p_square} in the temperature range of the current experiment. 

%

\begin{widetext}

\subsection{Appendix: Expressions for $H_d$, $H$, $F_d$, $F$, $A$, and $B$}\label{appendix}

\begin{align}
H_d^{(i)} &= \frac{1}{2^3 \pi^{3/2}} \int{ d^2\hat{p} \frac{ \hat{p}_i^2 }{ \left( \sum_{j=\{ x,y,z \}}{ \hat{p}_j^2 \frac{ \bar{\omega}^2 }{ \omega_j^2 }} \right)^{3/2} } \left[ (\hat{p} \cdot \hat{\epsilon})^2 - \frac{1}{3} \right]  }
\end{align}

\begin{align}
H^{(i)} &= \frac{1}{2^4 \pi^{3/2}} \int{ d^2\hat{p} \frac{ \hat{p}_i^2 }{ \left( \sum_{j=\{ x,y,z \}}{ \hat{p}_j^2 \frac{ \bar{\omega}^2 }{ \omega_j^2 }} \right)^{3/2} }  }
\end{align}

\begin{align}
F_d^{(i)} &= \frac{1}{2^3 \pi^{3/2}} \int{ d^2\hat{p} \frac{ \sqrt{ \left( \sum_{j=\{ x,y,z \}}{\frac{\omega_j^2 \hat{p}_j^2 }{\omega_i^2}} \right) -1 } - \arccos{\left[ \sqrt{ \frac{ \omega_i^2 }{ \sum_{j=\{ x,y,z \}}{ \hat{p}_j^2 \omega_j^2 } } } \right]} }{ \left[ \left( \sum_{j=\{ x,y,z \}}{ \frac{ \omega_j^2 \hat{p}_j^2 }{ \omega_i^2 } } \right) -1 \right]^{3/2} } \left[ (\hat{p} \cdot \hat{\epsilon} )^2 - \frac{1}{3} \right] }
\end{align}

\begin{align}
F^{(i)} &= \frac{1}{2^4 \pi^{3/2}} \int{ d^2\hat{p} \frac{ \sqrt{ \left( \sum_{j=\{ x,y,z \}}{\frac{\omega_j^2 \hat{p}_j^2 }{\omega_i^2}} \right) -1 } - \arccos{\left[ \sqrt{ \frac{ \omega_i^2 }{ \sum_{j=\{ x,y,z \}}{ \hat{p}_j^2 \omega_j^2 } } } \right]} }{ \left[ \left( \sum_{j=\{ x,y,z \}}{ \frac{ \omega_j^2 \hat{p}_j^2 }{ \omega_i^2 } } \right) -1 \right]^{3/2} } } 
\end{align}

\begin{align}
A_0^{(i)} &= \frac{3}{2^6 \pi^2} \int{d^2\hat{p} \int{ d^2\hat{p}' \frac{ ( \hat{p}_i'^2 - \hat{p}_i^2 ) \left[ \frac{ ( \hat{p}' \cdot \hat{\epsilon} )^2 + ( \hat{p} \cdot \hat{\epsilon} )^2 - 2 (\hat{p}' \cdot \hat{\epsilon}) (\hat{p} \cdot \hat{\epsilon}) (\hat{p}' \cdot \hat{p})}{ 1- (\hat{p} \cdot \hat{p}')^2} - \frac{2}{3} \right]^2 }{ \sqrt{ \sum_{j=\{ x,y,z \}}{\hat{p}^2_j \frac{\omega_j^2}{\bar{\omega}^2}} } } }}
\end{align}

\begin{align}
A_1^{(i)} &= \frac{3}{2^5 \pi^2} \int{d^2\hat{p} \int{ d^2\hat{p}' \frac{ ( \hat{p}_i'^2 - \hat{p}_i^2 ) \left[ \frac{ ( \hat{p}' \cdot \hat{\epsilon} )^2 + ( \hat{p} \cdot \hat{\epsilon} )^2 - 2 (\hat{p}' \cdot \hat{\epsilon}) (\hat{p} \cdot \hat{\epsilon}) (\hat{p}' \cdot \hat{p})}{ 1- (\hat{p} \cdot \hat{p}')^2} - \frac{2}{3} \right] }{ \sqrt{ \sum_{j=\{ x,y,z \}}{\hat{p}^2_j \frac{\omega_j^2}{\bar{\omega}^2}} } } }}
\end{align}

\begin{align}
A_2^{(i)} &= \frac{3}{2^6 \pi^2} \int{d^2\hat{p} \int{ d^2\hat{p}' \frac{ ( \hat{p}_i'^2 - \hat{p}_i^2 ) }{ \sqrt{ \sum_{j=\{ x,y,z \}}{\hat{p}^2_j \frac{\omega_j^2}{\bar{\omega}^2}} } } }}
\end{align}

\begin{align}
B_0^{(i)} &= \frac{3}{2^7 \pi^2} \int{d^2\hat{p} \int{ d^2\hat{p}' \frac{ ( \hat{p}_i'^2 - \hat{p}_i^2 ) \left[ \frac{ ( \hat{p}' \cdot \hat{\epsilon} )^2 + ( \hat{p} \cdot \hat{\epsilon} )^2 - 2 (\hat{p}' \cdot \hat{\epsilon}) (\hat{p} \cdot \hat{\epsilon}) (\hat{p}' \cdot \hat{p})}{ 1- (\hat{p} \cdot \hat{p}')^2} - \frac{2}{3} \right]^2 }{ \sqrt{ \sum_{j=\{ x,y,z \}}{ (3 \hat{p}^2_j+\hat{p}_j'^2) \frac{\omega_j^2}{\bar{\omega}^2}} } } }}
\end{align}

\begin{align}
B_1^{(i)} &= \frac{3}{2^6 \pi^2} \int{d^2\hat{p} \int{ d^2\hat{p}' \frac{ ( \hat{p}_i'^2 - \hat{p}_i^2 ) \left[ \frac{ ( \hat{p}' \cdot \hat{\epsilon} )^2 + ( \hat{p} \cdot \hat{\epsilon} )^2 - 2 (\hat{p}' \cdot \hat{\epsilon}) (\hat{p} \cdot \hat{\epsilon}) (\hat{p}' \cdot \hat{p})}{ 1- (\hat{p} \cdot \hat{p}')^2} - \frac{2}{3} \right] }{ \sqrt{ \sum_{j=\{ x,y,z \}}{ (3 \hat{p}^2_j+\hat{p}_j'^2) \frac{\omega_j^2}{\bar{\omega}^2}} } } }}
\end{align}

\begin{align}
B_2^{(i)} &= \frac{3}{2^7 \pi^2} \int{d^2\hat{p} \int{ d^2\hat{p}' \frac{ ( \hat{p}_i'^2 - \hat{p}_i^2 ) }{ \sqrt{ \sum_{j=\{ x,y,z \}}{ (3 \hat{p}^2_j+\hat{p}_j'^2) \frac{\omega_j^2}{\bar{\omega}^2}} } } }}
\end{align}
Here we used the notation $\hat{p}=(\sin{\theta}\cos{\phi},\sin{\theta}\sin{\phi},\cos{\theta})$, and $\int{d^2 \hat{p}}=\int_0^\pi \sin\theta d\theta\int_0^{2\pi}d\phi$.
\end{widetext}


\begin{thebibliography}{41}
\bibitem [{bli()}]{blindspot:footnote}%
  \BibitemOpen
  \href@noop {} {}\bibinfo {note} {Atom loss is significant near the resonance
  center, making the blind spot region in scattering length even
  larger.}\BibitemShut {Stop}%
\end{thebibliography}
\end{document}